\documentclass[12pt]{article}
\usepackage{amssymb}
\usepackage{amsmath}

\newtheorem{theorem}{Theorem}

\newtheorem{corollary}[theorem]{Corollary}

\newtheorem{definition}[theorem]{Definition}

\begin{document}
	
	\title{Lie's derivative in fluid mechanics}

	\author{ Henri Gouin \thanks{   
			E-mails:
			henri.gouin@univ-amu.fr; henri.gouin@yahoo.fr} }
	\date{\footnotesize Aix--Marseille University, CNRS,
		IUSTI, UMR 7343, Marseille, France.}
	\maketitle

	\date{}
	
	\maketitle

	\begin{abstract}

		The invariance theorems obtained in analytical mechanics and derived from Noether's theorems can be adapted to fluid mechanics. For this purpose, it is useful to give a functional representation of the fluid motion and to interpret the invariance group with respect to time in the quadri--dimensional reference space of Lagrangian variables. A powerful method of calculation uses  Lie's derivative, and many invariance theorems and conservation laws can be obtained in fluid mechanics. \\
		
		 
	\end{abstract}

	\section{Introduction}

	In analytical mechanics of systems with a finite number of degrees of freedom, Noether's theorem  is expressed as follows \cite{Gouin1}:\\
	We consider a differentiable curve $(C)$ of $\mathbb R^n$ and a form field in the dual $ R^{n\star}$ of  $\mathbb R^{n}$, $\boldsymbol{M}\in \mathbb R^n    \longrightarrow\boldsymbol \varXi^T(\boldsymbol{M})\in \mathbb R^{n\star}$. Then   $$a=\int_{(C)} \boldsymbol \varXi^T(\boldsymbol{M}) \,d\boldsymbol{M}$$ is a functional of $(C)$ denoted $a=\mathcal G(C)$. We can look for  paths such that $a$ is   extremal.\\
	We assume that $\displaystyle \left\{\mathcal T_u\right\},\, u\in I$, 
	where $I$ is an open set containing $0$, is a Lie group generating the solutions of autonomous differential equation  \begin{equation*}
	\frac{d\boldsymbol{M}}{du}= G(\boldsymbol{M}), 
	\end{equation*} 
	where the vector field $G(\boldsymbol{M})$  is the infinitesimal operator of the Lie group.\\ We call $\mathcal T_u(C)$ the transformed curve of $(C)$ by $\mathcal T_u$.
	If $a$ is invariant by  $\mathcal T_u$ \big (i.e.
	$
	\forall u\in I,\, \mathcal G\left(T_u(C)\right) =\mathcal G\left( C\right) 
	$\big), along extremal of $a$
	we obtain the relation   
	\begin{equation}
	\boldsymbol \varXi^T(\boldsymbol{M})\, G(\boldsymbol{M})=c\,,\label{Noethertheorem}
	\end{equation}
	where $c$ is  constant. Relation \eqref{Noethertheorem} is the analytical form of   Noether's theorem.\\
	
	\noindent We  show that for  fluid media, the velocity field in space--time plays a role analogous to the vector field $G(\boldsymbol{M})$ and    Hamilton's action states   theorems of invariance allowing the search of first integrals of motion \cite{Gavrilyuk1}; this is Noether's theorem generalized to fluids \cite{Gouin2}. Since  continuous media cannot be represented by a finite number of parameters, the forms of the first integrals are expressed   as conservation laws \cite{Dafermos}.  
	The natural method related to the definition of   motions of a fluid uses a group with one--time parameter corresponding to a diffeomorphism representing the   fluid motion at each instant. The velocity field in space--time is the infinitesimal generator of this group \cite{Casal}. Its main advantage is  to use the variance of specific medium quantities which have the structure of scalars, vectors, forms and tensors  covariant order $p$ and  contravariant order $q$. When they do not depend on the trajectories, the quantities  can be defined on the reference space independently of the motions \cite{Serrin}. They are related to the Lie derivative associated with the velocity field  in the time--space of the fluid motion \cite{yano}. Their variance allows us to  use   the linear tangent application associated with the  motion.\\
	Conservative laws are well studied in the literature \cite{Benjamin,Yahalow,Shankar,Serre,Kambe}. In our article,
	we obtain conservation laws in a geometrical form, the simplest forms are associated with scalar fields allowing to write conservation laws in a time-space divergence form. We analyse the different tensors classically obtained in the literature and we can obtain first integrals of motions. \\    
	
	\noindent\emph{Notations:} For any vectors $\boldsymbol{a,b}$, term $\boldsymbol{a}%
	^{T }\boldsymbol{b}$ denotes the scalar product (line vector
	$\boldsymbol{a}^{T}$
	is multiplied by column vector $\boldsymbol{b}$) and tensor $\boldsymbol{a} {%
		\ }\boldsymbol{b}^{T }$ (or $\boldsymbol{a}\otimes \boldsymbol{b}$) denotes
	the product of column vector $\boldsymbol{a}$ by line vector $\boldsymbol{b}%
	^{T},$ where superscript $^T$ denotes the transposition. Tensor $\boldsymbol{%
		I}$ denotes the identity transformation.\\
	The gradient of  scalar function $f(\boldsymbol x)$ is the transposition $\displaystyle\left(\frac{\partial f}{\partial
		{\boldsymbol x}}\right)^T$ associated with  linear form $df=\displaystyle  \displaystyle\left(\frac{\partial f}{\partial
		{\boldsymbol x}}\right)\,d{\boldsymbol x}$  and  $\displaystyle\frac{\partial \boldsymbol v(\boldsymbol x)}{\partial
		{\boldsymbol x}}$\  denotes the linear application defined by the relation $\displaystyle d\boldsymbol v=   \left(\frac{\partial \boldsymbol v(\boldsymbol x)}{\partial
		{\boldsymbol x}}\right)\, d\boldsymbol x$.

	\section{Lie's derivative associated with   a fluid motion}
	
	\subsection{Generalities}
	The  fluid  motion  is represented by a   $t$-dependent $C^{2}$--diffeomorphism $\varphi _{t}$ \cite{Gouin3}:
	\begin{equation*}
	\boldsymbol{X} \in \mathcal{D}_{0}\ \longrightarrow\ \boldsymbol{x} \in \mathcal{D}%
	_{t}\,     ,
	\end{equation*}
	such  that   $
	\boldsymbol{x}=\boldsymbol{\varphi}_t(\boldsymbol{X}) $  is a mapping
	from a reference space ${\mathcal D}_{0}$ of Lagrange variables $ \boldsymbol{X}= (X_1,X_2,X_3)^T$ into  the space ${\mathcal D}_{t}$ of Euler variables $ \boldsymbol{x}= (x_1,x_2,x_3)^T$ occupied by the fluid at time $t$.\\ 
	
	\noindent The linear tangent mapping of $\boldsymbol{\varphi}_t$ is defined by: 
	\begin{equation*}
	\boldsymbol F=\frac{\partial \boldsymbol x}{\partial \boldsymbol X}\equiv  \frac{\partial \boldsymbol{\varphi}_t(\boldsymbol{X}) }{\partial \boldsymbol X}.
	\end{equation*}
	By derivation with respect to time, we get:
	\begin{equation}
	\frac{d\boldsymbol F}{dt}=\displaystyle\frac{ \partial \boldsymbol{u}%
	}{\partial \boldsymbol{x}}\, \boldsymbol F\quad   {\rm and}\quad  \displaystyle\frac{d\boldsymbol F^{-1}}{dt}=-\boldsymbol F^{-1}%
	\frac{\partial \boldsymbol{u}}{\partial \boldsymbol{x}}, \label{deriv}
	\end{equation} 
	where  $\boldsymbol{u}$  is the velocity   in ${\mathcal D}_{t}$.\\
	The motion of the fluid can be also represented by the inversible  differentiable  mapping ${\boldsymbol\Phi}$ such that \cite{Casal,Gouin3}:  
	\begin{equation*}
	\boldsymbol{z}=\boldsymbol{\Phi} (\boldsymbol{Z}) \quad \text {with}\quad\boldsymbol{z}=  %
	\left[\begin{array}{c}
	t\\
	\boldsymbol{x}
	\end{array}\right]\equiv  %
	\left[\begin{array}{c}
	t\\
	\boldsymbol{\varphi}_t(\boldsymbol{X})
	\end{array}\right]
	\in {\mathcal W}\quad \text{and}\quad  \boldsymbol{Z}= \left[\begin{array}{c}
	t\\
	\boldsymbol{X}
	\end{array}\right]\in {\mathcal W}_{0}, 
	\end{equation*} 
	where ${\mathcal W}$ is the  \textit{4-D} physical space-time and  ${\mathcal W}_0$ is the \textit{4-D}  reference space. 
	In  differential manifold ${\mathcal W}$, the space-time velocity is  $\mathcal U=\displaystyle \left[\begin{array}{c}
	{\it{1}}\\
	\boldsymbol{u}
	\end{array}\right]$. \\ 
	In each point of  ${\mathcal W}$, the \textit{4-D}  vector $\mathcal U$ generates  a local one--parameter group defined by the mapping 
	\begin{equation*}
	(h, \boldsymbol{z}) \in\,]-\epsilon,\,\epsilon[\,\times\, \mathcal O\longrightarrow\, \Theta(h,  \boldsymbol{z})\in   {\mathcal W} ,
	\end{equation*}
	such that
	\begin{equation*}
	\Theta(h,  \boldsymbol{z}) =\left[\begin{array}{c}
	h+t\\ \boldsymbol{\varphi}_{h+t}
	\left(\boldsymbol{\varphi}_t^{-1}
	(\boldsymbol{x})\right)
	\end{array}\right]\equiv \left[\begin{array}{c}
	h+t\\ \boldsymbol{\varphi}_{h+t}
	\left( 
	\boldsymbol{X}\right)
	\end{array}\right],
	\end{equation*}
	where $\mathcal O$ is an open set of  $ {\mathcal W}$ and $\epsilon$ a positive real number.  The velocity field $\mathcal {U}$ is the infinitesimal
	displacement of the one-parameter group associated with transformation $\boldsymbol{\Phi}$. 
	\\
	We consider  tensor fields  with covariant order $p$ and contravariant order $q$   applied to the vectors of the tangent space to $\mathcal D_t$ and   the forms  of its dual (or covector of cotangent space) at $\boldsymbol x$, respectively; we said tensors in vector  space $T_{\boldsymbol x}^{p\, q\star}(\mathcal D_t)$. At each point $\boldsymbol x$ of $\mathcal D_t$ we consider the fiber-space (fiber-bundle)  which is  tangent  order $p$ and    cotangent order $q$\ \cite{Steenrod}.\\
	To each tensor field $\boldsymbol v \in T_{\varphi_t(\boldsymbol X)}^{p\, q\star}(\mathcal D_t)$ when   $\boldsymbol z   \in {\mathcal O}$,  we associate  its image      in the tensor space $T_{ \boldsymbol{\varphi}_{h+t}	
		(\boldsymbol{X})}^{p \, q\star}(\mathcal D_{t+h})$  when $\Theta(h,  \boldsymbol{z}) \in {\mathcal O} $. We   denote the image: 
	\begin{equation*}
	\boldsymbol v_h = \theta_h(\boldsymbol v) .
	\end{equation*}
	\begin{definition}
		\textit{A tensor field is moving with the fluid if and only if its image in the reference space represented in Lagrange variables is independent of the time.} 
	\end{definition}
	The definition is an extension of set $\mathcal E_0$ of $\mathcal D_0$ and its image $\mathcal E$:
	\begin{equation*}
	\mathcal E = \left\{\boldsymbol x\,\ {\rm such\ that}\quad\boldsymbol x =\varphi(\boldsymbol X),\ \boldsymbol X \in \mathcal E_0\right\},
	\end{equation*}
	when   time $t$ varies. The set $\mathcal E$ is always made of the same particles. The set $\mathcal E$ is material or moving with the fluid. This is in particular the case of curves, surfaces and volumes of     $\mathcal D_0$.
	\\
	
	\noindent To this definition, we associate a special derivative of the tensor fields.
	\begin{definition}
		\textit{The Lie derivative of $\boldsymbol v$  is   $\displaystyle  d_L \boldsymbol v= \lim_{h\rightarrow 0}\left( \frac{\theta_h(\boldsymbol v(\boldsymbol{z}))- \boldsymbol v(\boldsymbol{z})}{h}\right)$.}
	\end{definition}
	To interpret $d_L\boldsymbol v$ in fluid mechanics, ${\mathcal W}_0$ is a convenient space. Let $\boldsymbol v_0(\boldsymbol Z)$ the image of  $\boldsymbol v(\boldsymbol z)$ by  ${\boldsymbol{\tilde\Psi}}$ the mapping for the tensor field corresponding to $\boldsymbol{\Psi}=\boldsymbol{\Phi}^{-1}$  and we denote ${\boldsymbol{\tilde\Phi}}$ the mapping corresponding to  ${\boldsymbol{\Phi}}$. We obtain  
	\begin{equation*} 
	\displaystyle d_L \boldsymbol v =\displaystyle\lim_{h\rightarrow 0}\left( \frac{\tilde{\boldsymbol \Phi}\left(\tilde{\boldsymbol \Psi}\,\left(\theta_h(\boldsymbol v(\boldsymbol{z}))\right)-\tilde{\boldsymbol \Psi}(\boldsymbol v(\boldsymbol{z}))\right)}{h} \right)
	\end{equation*}
	\begin{equation*}
	\qquad = \displaystyle \lim_{h\rightarrow 0}\tilde{\boldsymbol \Phi}\left( \frac{\boldsymbol v_0(t+h,\boldsymbol{X}) -  \boldsymbol v_0(t,\boldsymbol{X})}{h} \right).
	\end{equation*} 
	Then,
	\begin{equation*}
	d_L \boldsymbol v =  \tilde{\boldsymbol \Phi}  \left(\frac{d\boldsymbol v_0(t,\boldsymbol{X})}{dt}\right)
	\end{equation*}
	corresponds to the commutative graph:  
	\begin{equation*}
	\begin{array}{ccc}
	\boldsymbol v\in T_{\boldsymbol x}^{p, q\star}(\mathcal D_t) &
	\begin{array}{c}
	\tilde{\boldsymbol \Psi} \\
	\quad	\longrightarrow\quad \\
	\\
	\end{array}
	& \boldsymbol v_0 \in T_{\boldsymbol X}^{p, q\star}(\mathcal D_0) \\
	\begin{array}{c}
	\displaystyle d_{L}\,\Big\downarrow \\
	\end{array}
	&  &
	\begin{array}{c}
	\displaystyle\Big\downarrow\, \displaystyle\frac{d}{dt} \\
	\end{array}
	\\
	\begin{array}{c}
	\\
	\displaystyle d_L \boldsymbol v \in T_{\boldsymbol x}^{p, q\star}(\mathcal D_t)
	\end{array}%
	&
	\begin{array}{c}
	\tilde{\boldsymbol\Phi}\\
	\longleftarrow \\
	\end{array}
	&
	\begin{array}{c}
	\\
	\displaystyle\frac{d\boldsymbol v_0(t,\boldsymbol X) }{dt}   \in T_{\boldsymbol X}^{p, q\star}(\mathcal D_0) 
	\end{array}%
	\end{array}
	\label{graph}
	\end{equation*}
	To consider tensor fields,  we can also consider  restrictions of tensors  defined   on ${\mathcal W}$ at $t$ given (the Lie derivative does not commute with the restriction).
	\\
	To consider tensor fields, we can represent their images in  ${\mathcal W}_0$ and we assume that the images depends only on $\boldsymbol{X}$ and do  not depend on the trajectory parameter  (i.e.  time). Such  quantities are called \textit{moving with the fluid} and its definition domain is ${\mathcal D}_0$.
	From the previous results, we can write:  
	\begin{theorem}
		A tensor is moving with the fluid if and only if its Lie derivative associated with the vector field\,\ ${\mathcal U}$ is zero.
	\end{theorem}
	\subsection{Lie's derivative of  tensor fields}
	\subsubsection{Scalar field}
	A scalar  field is moving with the fluid if and only  its Lie's derivative is null (which corrresponds to its material derivative) 
	\begin{equation*}
	\exists\ \{\boldsymbol X\in \mathcal {D}_0\longrightarrow s_0(\boldsymbol X)\}\quad\text{such that}\quad s(t,\boldsymbol x)= s_0(\boldsymbol X).
	\end{equation*}
	For example,
	in isentropic (conservative) motions, the specific entropy is moving with the fluid: 
	\begin{equation*}
	\frac{ds} {dt} \equiv\frac{\partial s}{\partial t}+\frac{\partial s}{\partial \boldsymbol{x}}\,\boldsymbol{u}=0,
	\end{equation*}
	where the  $d/dt$ means the material  derivative.
	\subsubsection{Vector field}
	The tangent linear mapping $\boldsymbol F =\dfrac{\partial\boldsymbol{x}}{\partial\boldsymbol{X}}$ transforms  vectors of the tangent vector space $T_{\boldsymbol{X}}^1(\mathcal D_0)$ at $\boldsymbol{X}\in
	\mathcal D_0$  into vectors of the tangent vector space $T_{\boldsymbol{x}}^1(\mathcal D_t)$ at $\boldsymbol{x}\in
	\mathcal D_t$.\\
	Let us define a vector field $\boldsymbol J$ of $\mathcal{D}_{t}$  by the mapping:
	\begin{equation*}
	\boldsymbol{z} \in W  \ \longrightarrow\ \boldsymbol J(\boldsymbol{x},t)\in T_{%
		\boldsymbol{x}}^1(\mathcal{D}_{t}).
	\end{equation*}%
	The
	Lie derivative $d_{L}$ of the vector field $\boldsymbol J$ is deduced from the commutative
	diagram:\\
	\begin{equation*}
	\begin{array}{ccc}
	\boldsymbol J\in T_{\boldsymbol x}  ^1    (\mathcal{D}_{t}) &
	\begin{array}{c}
	\displaystyle\tilde{\boldsymbol\Psi}   \\
	\longrightarrow \\
	\\
	\end{array}
	&\boldsymbol F^{-1} \boldsymbol J\,\in T_{\boldsymbol X}^1(\mathcal{D}_{o}) \\
	\begin{array}{c}
	\displaystyle d_{L}\,\Big\downarrow \\
	\end{array}
	&  &
	\begin{array}{c}
	\displaystyle\Big\downarrow\, \displaystyle\frac{d}{dt} \\
	\end{array}
	\\
	\begin{array}{c}
	\\
	\displaystyle	\frac{d{\boldsymbol{J}}}{dt}-\dfrac{\partial\boldsymbol{v}}{\partial\boldsymbol{x}}\,{\boldsymbol{J}}\in T_{\boldsymbol x}^1 (\mathcal{D}_t) 
	\end{array}%
	\ \ \ \ \ \  &
	\begin{array}{c}
	\displaystyle\displaystyle\tilde{\boldsymbol\Phi}  \\
	\longleftarrow \\
	\end{array}
	&
	\begin{array}{c}
	\ \ \ \ \ \ \ \  \\
	\ \ \ \displaystyle\frac{d{\boldsymbol{F}^{-1}}}{dt}\boldsymbol{J}  + \boldsymbol{F}^{-1}\ {\frac{d{\boldsymbol{J}}}{dt}}\in T_{\boldsymbol{X}}^1(\mathcal{D}_0)%
	\end{array}%
	\end{array}
	\label{graph}
	\end{equation*}
	\bigskip
	
	\noindent 
	Consequently, the Lie derivative of $\boldsymbol{J}$ is:
	$$d_L\boldsymbol{J}  ={\frac{d{\boldsymbol{J}}}{dt}}-\dfrac{\partial\boldsymbol{u}}{\partial\boldsymbol{x}}{\boldsymbol{J}}\equiv\frac{\partial \boldsymbol{J}}{\partial t}+\dfrac{\partial\boldsymbol J}{\partial\boldsymbol{x}}{\boldsymbol{u}}-\dfrac{\partial\boldsymbol u}{\partial\boldsymbol{x}}{\boldsymbol{J}}$$ 
	Let us note that
	$\displaystyle\frac{d\boldsymbol J}{dt}-\displaystyle\frac{\partial
		\boldsymbol{u}}{\partial \boldsymbol{x}}\, \boldsymbol J\,$ is the convective derivative of $\boldsymbol J$ with
	respect to the velocity field $\mathcal {U}$. 
	\begin{theorem}
		$d_L\boldsymbol{J} =\boldsymbol{0}$\,\ if and only if 
		\begin{equation*}
		\exists\ \{ \boldsymbol X\in \mathcal {D}_0\longrightarrow {\boldsymbol{J}_0}(\boldsymbol X)\in T_{\boldsymbol{X}}^1(\mathcal{D}_0)\}\quad {\rm such\ that}\quad \boldsymbol{J}(t, \boldsymbol x)= \boldsymbol{F}\boldsymbol{J}_0(\boldsymbol X)
		\end{equation*}
	\end{theorem}
	
	\subsubsection{Form field}
	Let us denote $T^{1\ast }_{\boldsymbol x}(\mathcal{D}_{t})$ the cotangent linear space of $%
	\mathcal{D}_{t}$ at $\boldsymbol x$ and $T_{\boldsymbol{X}}^{1\ast }(\mathcal{D}_{o})$ the cotangent
	linear space of $\mathcal{D}_{o}$ at $\boldsymbol{X}$, and consider a field $\boldsymbol C$ of  forms (covectors) of $\mathcal{D}_{t}$:
	\begin{equation*}
	\boldsymbol{z} \in \mathcal W \longrightarrow\ \boldsymbol C(\boldsymbol{x},t)\in T_{%
		\boldsymbol{x}}^{1\ast }(\mathcal{D}_{t}).
	\end{equation*}%
	The
	Lie derivative $d_{L}$ of the form field $\boldsymbol C$ is deduced from the commutative
	diagram:
	\begin{equation*}
	\begin{array}{ccc}
	\boldsymbol C\in T^{1\ast }_{\boldsymbol x}      (\mathcal{D}_{t}) &
	\begin{array}{c}
	\displaystyle\tilde{\boldsymbol\Psi}  \\
	\longrightarrow \\
	\\
	\end{array}
	& \boldsymbol C\,\boldsymbol F\in T^{1\ast }_{\boldsymbol X}(\mathcal{D}_{o}) \\
	\begin{array}{c}
	\displaystyle d_{L}\,\Big\downarrow \\
	\end{array}
	&  &
	\begin{array}{c}
	\displaystyle\Big\downarrow\, \displaystyle\frac{d}{dt} \\
	\end{array}
	\\
	\begin{array}{c}
	\\
	\displaystyle\frac{d\boldsymbol C}{dt}+\boldsymbol C\,\displaystyle\frac{\partial \boldsymbol{u}}{\partial
		\boldsymbol	x}\in T^{1\ast }_{\boldsymbol x}(\mathcal{D}_{t})
	\end{array}%
	\ \ \ \ \ \  &
	\begin{array}{c}
	\displaystyle\tilde{\boldsymbol\Phi}  \\
	\longleftarrow \\
	\end{array}
	&
	\begin{array}{c}
	\ \ \ \ \ \ \ \  \\
	\ \ \ \displaystyle\frac{d\boldsymbol C}{dt}\boldsymbol F+\boldsymbol C\, \displaystyle\frac{\partial \boldsymbol{u}}{%
		\partial \boldsymbol x} \boldsymbol F\in T^{1\ast }_{\boldsymbol X}(\mathcal{D}_{o})%
	\end{array}%
	\end{array}
	\label{graph}
	\end{equation*}
	\bigskip
	
	\noindent 
	Consequently, the Lie derivative of $\boldsymbol{C}$ is:
	\begin{equation*}
	d_L \boldsymbol C =  {\frac{{d \boldsymbol C}}{dt}}+\boldsymbol C \,  \frac{\partial \boldsymbol u}{\partial \boldsymbol{x}}\equiv\frac{\partial \boldsymbol{C}}{\partial t}+\boldsymbol u^T\left(\frac{\partial \boldsymbol C^T}{\partial \boldsymbol{x}}\right)^T +\boldsymbol C \,  \frac{\partial \boldsymbol u}{\partial \boldsymbol{x}}.
	\end{equation*}
	
	\begin{theorem}
		$d_L \boldsymbol C  =\boldsymbol{0}$\;\ if and only if 
		\begin{equation*}
		\exists\ \{\boldsymbol X\in \mathcal {D}_0\longrightarrow {\boldsymbol{C}_0}(\boldsymbol X)\in T_{\boldsymbol{X}}^{1\star} (\mathcal{D}_0)\}\quad{\rm such\ that}\quad \boldsymbol{C}(t, \boldsymbol x)= \boldsymbol{C}_0(\boldsymbol X)\boldsymbol{F}^{-1} .
		\end{equation*} 
	\end{theorem} 
	\bigskip
	For all form field\ 
	$
	\boldsymbol{z} \in W\ \longrightarrow\ \boldsymbol C(t, \boldsymbol{x})\in T_{\boldsymbol{x}}^{1\star} (\mathcal{D}_t) 
	$ and for all curve $(\gamma)$ in $\mathcal{D}_t$, we associate the curvilinear integral
	$I = \displaystyle\int_\gamma \boldsymbol C(t, \boldsymbol{x})\,d\boldsymbol{x}$.
	\begin{corollary} 
		A form field is moving with the fluid if and only is, for all  transported vector field, the associated scalar product is a transported scalar field.
	\end{corollary}
	\begin{theorem}
		Form $\boldsymbol C$  being a transported form field, its integral along any fluid curve is constant. If we denote by $\gamma_0$ the image of $\gamma$ in $\mathcal{D}_0$, we obtain 
		\begin{equation*}
		\int_\gamma C(t, \boldsymbol{x})\,d\boldsymbol{x} = 	\int_{\gamma_0} C_0(\boldsymbol{X})\,d\boldsymbol{X}.
		\end{equation*} 
	\end{theorem}
	This result leads to Kelvin's theorems.
	
	\subsubsection{2--form field}
	We consider a 2--form field  $\boldsymbol{z} \in \mathcal W\ \longrightarrow\ \omega(t, \boldsymbol{x})\in T_{\boldsymbol{x}}^{2\star}(\mathcal{D}_t) $. There exists an isomorphism $\omega(t, \boldsymbol{x})\in T_{\boldsymbol{x}}^{2\star}(\mathcal{D}_t) \longrightarrow \boldsymbol {W}\in
	T_{\boldsymbol{x}}^1(\mathcal{D}_t) $ defined as  
	\begin{equation}
	\forall\ \boldsymbol{v_1}\ \text{and}\  \boldsymbol{v_2}\in
	T_{\boldsymbol{x}}^1(\mathcal{D}_t),\quad \omega\, (\boldsymbol{v_1}, \boldsymbol{v_2})=\text{det}\,(\boldsymbol{W}, \boldsymbol{v_1}, \boldsymbol{v_2}),\label{2form}
	\end{equation}
	where\,\ "det"\,\ is the application determinant. We are back to the tensorial structure of $\boldsymbol{W}$ that does not have the structure of a vector field. Vector fields $\boldsymbol{v_1}$ and  $\boldsymbol{v_2}$ defined in 
	$T_{\boldsymbol{x}}^1 (\mathcal{D}_t) $ have images 
	$\boldsymbol{v_{10}}$ and  $\boldsymbol{v_{20}}$  in 
	${T_{\boldsymbol{X}}^1}(\mathcal{D}_0) $. From \eqref{2form} we can write:  
	\begin{equation*}
	\text{det}\,\left(\boldsymbol{W}, \boldsymbol{v_1}, \boldsymbol{v_2}\right) = \text{det}\,(\boldsymbol{W_0}, \boldsymbol{v_{10}}, \boldsymbol{v_{20}}).
	\end{equation*}
	Then,
	\begin{equation*}
	\boldsymbol{W} \,(t, \boldsymbol{x}) =\frac{\boldsymbol{F}\,\boldsymbol{W_0\,(t, \boldsymbol{X})}}{\text{det}\,\boldsymbol{F}} 
	\end{equation*}
	and we deduce  immediately:
	\begin{equation*}
	d_L\big(\text{det}\,({\boldsymbol{W},.,.)}\big)= \text{det}\,\left(\frac{d\boldsymbol{ W}}{dt}+\boldsymbol{W}\, \text{div}\,\boldsymbol u -\frac{\partial\boldsymbol u}{\partial\boldsymbol{x}}\, \boldsymbol{W},.,.\right),
	\end{equation*}
	where\,\ div\,\ is the application divergence in $\mathcal D_t$.
	\begin{corollary} 
		A 2-form field is moving with the fluid if and only if,  applied to  any two transported vector fields, we get a   transported scalar field.
	\end{corollary}
	From theses results we define the integral of a 2-form field on a surface $S$ of $\mathcal{D}_t$. We denote $S_0$ the image of $S$ in $\mathcal{D}_0$; we obtain  
	\begin{equation*}
	\iint_S \text{det}\,\left(\boldsymbol{W}, d_1\boldsymbol{x},d_2\boldsymbol{x}\right)=\iint_{S_0} \text{det}\,\left(\boldsymbol{W_0}, d_1\boldsymbol{X},d_2\boldsymbol{X}\right)
	\end{equation*}
	This integral corresponds to the flux of $\boldsymbol W$ through $S$. We obtain the following properties :
	\begin{theorem}
		The five propositions are equivalent : \\
		
		-\quad The flux of $\boldsymbol{W}$ through any fluid surface is constant.\\
		
		-\quad	The 2-form field $\text{det}\,({\boldsymbol{W},.,.) }$ is moving with the fluid.\\
		
		-\quad The vector field $\boldsymbol W\, \text{det}\,\boldsymbol F$ is moving with the fluid.\\
		
		-\quad $
		\exists \,\boldsymbol X\in \mathcal {D}_0\longrightarrow {\boldsymbol{W}_0}(\boldsymbol X)\in T_{\boldsymbol{X}}^1  (\mathcal{D}_0)\,\ {\rm such\ that}\,\ \boldsymbol{W}(t, \boldsymbol x)=\displaystyle\frac{\boldsymbol{F}}{ {\rm det}\, \boldsymbol{F}}\, \boldsymbol{W}_0(\boldsymbol X). 
		$\\
		
		-\quad $\displaystyle\frac{d\boldsymbol{W}}{dt} +\boldsymbol{W}\,  {\rm div}\,\boldsymbol u -\frac{\partial\boldsymbol u}{\partial\boldsymbol{x}}\, \boldsymbol{W}\equiv \frac{\partial\boldsymbol{W} }{\partial t}+\frac{\partial\boldsymbol{W}}{\partial\boldsymbol{x}}\, \boldsymbol u+\boldsymbol{W}\,  {\rm div}\,\boldsymbol u -\frac{\partial\boldsymbol u}{\partial\boldsymbol{x}}\, \boldsymbol{W}=\boldsymbol{0}.$
	\end{theorem}
	In the case of barotopic flow,
	we denote the vorticity of the fluid by 
	\begin{equation*}
	\boldsymbol\omega = {\rm curl}\,\boldsymbol u
	\end{equation*}
	we obtain the Helmholtz equation \cite{Serrin}:
	\begin{equation}
	\frac{D}{Dt}\left(\frac{\boldsymbol \omega}{\rho}\right)=\frac{\partial \boldsymbol u}{\partial \boldsymbol x}\,\frac{\boldsymbol \omega}{\rho} \label{vorticity4} 
	\end{equation}
	In particular, it implies the well-known Kelvin theorem. The developed form is the Helmholtz equation : 
	\begin{equation}
	\frac{\partial \boldsymbol \omega}{\partial t}+\frac{\partial \boldsymbol \omega}{\partial \boldsymbol x} {\boldsymbol u}+{\boldsymbol \omega}\,{\rm div} \,{\boldsymbol u}-\frac{\partial \boldsymbol u}{\partial \boldsymbol x} {\boldsymbol \omega}=0  \label{vorticity}
	\end{equation} 
	and $\boldsymbol\omega$ is associated to a 2-form moving with the fluid.
	\subsubsection{3-form field}
	A 3-form  of  $T_{\boldsymbol{x}}^{3\star}(\mathcal{D}_t)$ is an element of a real vector space of one dimension corresponding to the value of an isomorphism between 3-forms and scalar fields. If the 3-forms is written\,\ $v\,\text{det}$, \,
	for any vectors $\boldsymbol{v}_1, \boldsymbol{v}_2, \boldsymbol{v}_3$ of $T_{\boldsymbol{x}}^1 (\mathcal{D}_t) $ with images $\boldsymbol{v}_{01}, \boldsymbol{v}_{02}, \boldsymbol{v}_{03}$ of $T_{\boldsymbol{X}}^1 (\mathcal{D}_0)$, it exists $v_0$ scalar field  of  ${\mathcal W}_0$ such that  \\
	\begin{equation*}
	v \,  \text{det}\left(\boldsymbol{v}_1, \boldsymbol{v}_2, \boldsymbol{v}_3\right)= v_0 \,  \text{det}\left(\boldsymbol{v}_{01}, \boldsymbol{v}_{02}, \boldsymbol{v}_{03}\right)
	\end{equation*}
	and for any volume $\Omega_0$ in $\mathcal{D}_0$ of image $\Omega$ in $\mathcal{D}_t$ 
	\begin{equation*}
	\iiint_\Omega v \,  \text{det}\left(\boldsymbol{dx}_1, \boldsymbol{dx}_2, \boldsymbol{dx}_3\right)= 
	\iiint_{\Omega_0} v_0 \,  \text{det}\left(\boldsymbol{dX}_1, \boldsymbol{dX}_2, \boldsymbol{dX}_3\right).
	\end{equation*}
	Then,  $ v(\boldsymbol{x},t) \,  \text{det} \,\boldsymbol F= v_0(\boldsymbol{X},t) $ and the Lie derivative of\,\ $v \,  \text{det}$\,\ is:
	\begin{equation*}
	d_L(\, v \,  \text{det})= \left(\frac{d v}{dt} + v\, \text{div}\, \boldsymbol{u} \right) \,  \text{det} \ .
	\end{equation*}
	\begin{corollary}
		The 3-form field\,\ $v\,\text{det}$\,\ is moving with the fluid if and only if, applied to  any three transported vector fields, we get a   transported scalar field.\\
		If\,\ $v\,\text{det}$\,\ is moving with the fluid, the volume integral is constant in the convected fluid volume.
	\end{corollary}
	This property is equivalent to 
	\begin{equation*}
	\frac{dv}{dt} + v\, \text{div}\, \boldsymbol u\equiv \frac{\partial v}{\partial t}+\text{div}(v\,\boldsymbol u)\equiv \text{Div}(v\,\mathcal U) =0,   
	\end{equation*}
	where\,\ Div\,\ the space--time divergence in $\mathcal{W}$, corresponds to the conservation of   density $v$.\\ To define the specific mass, we assume a given mass distribution $\rho_0(\boldsymbol{X})$ in $\mathcal{D}_0$. The mass density $\rho$ is given by the relationship 
	\begin{equation*}
	\rho\ \text{det}\, \boldsymbol F = \rho_0(\boldsymbol{X})   
	\end{equation*}
	and from Jacobi's derivation of det $\boldsymbol F$ and relation \eqref{deriv}
	\begin{equation*}
	\frac{d(\text{det}\, \boldsymbol F)}{dt}=(\text{det}\, \boldsymbol F)\,{\rm Tr}\left( \boldsymbol F^{-1}\frac{d \boldsymbol F }{dt}\right)= (\text{det}\, \boldsymbol F)\,{\rm Tr}\left(\frac{\partial \boldsymbol{u}}{\partial \boldsymbol{x}}\right),
	\end{equation*}
	where "Tr" is the trace operator. We deduce:
	\begin{equation*}
	\frac{d\rho}{dt} + \rho\, \text{div}\, \boldsymbol u\equiv \frac{\partial \rho}{\partial t}+\text{div}(\rho\,\boldsymbol u)\equiv \text{Div}(\rho\,\mathcal U) =0 . 
	\end{equation*}
	\subsubsection{A general tensor field}
	\begin{definition} 
		A   tensor field  with covariant order $p$ and contravariant order $q$  is moving with the fluid if and only if, applied to any $p$ vectors and $q$ forms moving with the fluid,  the associated scalar  is moving with the fluid.
	\end{definition}
	This property is equivalent to a zero Lie derivative of the tensor field.\\
	An example is  matrix field  $\boldsymbol M$. We obtain the immediate property 
	\begin{equation*}
	\boldsymbol M (\boldsymbol x,t) =\boldsymbol F\, \boldsymbol M_0(\boldsymbol X)\, \boldsymbol F^{-1} \quad \Longleftrightarrow\quad d_L\boldsymbol M= \frac{d\boldsymbol{M}}{dt}  + \boldsymbol M\,\frac{\partial \boldsymbol u }{\partial \boldsymbol x}- \frac{\partial \boldsymbol u }{\partial \boldsymbol x}\, \boldsymbol M,
	\end{equation*}
	where $\boldsymbol M_0(\boldsymbol X)$ is the image matrix defined on $\mathcal{D}_0$.
	\\
	We can summarize the results of   previous paragraphs in  form of a table:
	
	{\footnotesize
		\begin{center}
			\begin{tabular}{|c|c|c|}
				\hline
				&    &
				\\   
				\multicolumn{1}{|c|}{Tensors} & {Lie's derivatives} & {Tensors moving with the fluid} 
				\\
				& &
				\\
				\hline 
				& &
				\\
				\multicolumn{1}{|c|}{ Scalar $s$} &  {$   \displaystyle\frac{ds}{dt} $} & $s(\boldsymbol x,t)=s_0(\boldsymbol X) $
				\\
				& &
				\\
				Vector $\displaystyle\boldsymbol J$ & 	{$   \displaystyle\frac{d\boldsymbol J}{dt}- \displaystyle\frac{\partial\boldsymbol u}{\partial\boldsymbol x} \,\boldsymbol J$} & $\boldsymbol J(\boldsymbol x,t) = \boldsymbol F\, \boldsymbol J(\boldsymbol X)$   \\ 
				&  &
				\\ 
				\multicolumn{1}{|c|} {Form 
					$\boldsymbol C$}  &  $ \displaystyle\frac{d\boldsymbol C}{dt}+ \boldsymbol C\, \displaystyle\frac{\partial\boldsymbol u}{\partial\boldsymbol x}$& $\boldsymbol C(\boldsymbol x,t)\, \boldsymbol F=   \boldsymbol C(\boldsymbol X)$  
				\\ 
				& &
				\\
				2-form
				$\text{det}\,(\boldsymbol{W},.,.)$  & $  \text{det}\,\left({\displaystyle\frac{d\boldsymbol W}{dt} }+\boldsymbol{W}\, \text{div}\,\boldsymbol u -\displaystyle\frac{\partial\boldsymbol u}{\partial\boldsymbol{x}}\, \boldsymbol{W},.,.\right)$ &$\boldsymbol{W}(\boldsymbol{x},t) = \displaystyle\frac{\boldsymbol{F}}{{\rm det}\, \boldsymbol{F}}\, \boldsymbol{W}_0(\boldsymbol{X})$
				\\ 
				& &
				\\
				3--form $v\,$det  & $\displaystyle\left(\frac{dv}{dt}+ v\, div\,\boldsymbol u\right)$  det & $v(\boldsymbol x,t)\, det\, \boldsymbol F = v_0(\boldsymbol X) $
				\\
				& &
				\\
				Matrix
				$\boldsymbol M$  & $\displaystyle \frac{d\boldsymbol M}{dt}+ \boldsymbol M\,\displaystyle\frac{\partial\boldsymbol u}{\partial\boldsymbol{x}}- \displaystyle\frac{\partial\boldsymbol u}{\partial\boldsymbol{x}}\,\boldsymbol M $ & $\boldsymbol M= \boldsymbol F\, \boldsymbol M_0(\boldsymbol X)\, \boldsymbol F^{-1}$
				\\
				& &
				\\
				\hline     
				
			\end{tabular}\vskip0.5cm
			
		\end{center}
	}

	\subsection{Lie's derivative and exterior derivative}
	It is possible to summarize the properties of integrals by noting the integral $\displaystyle \int_\Omega\pi\, d\omega$ of a $p$--form $\pi$ in a fluid domain $\Omega$ convected by the flow:
	\begin{theorem}
		a $p$--form $\pi$ in a fluid domain $\Omega$ convected by the flow	verifies
		\begin{equation*}
		\frac{d}{dt}\left(\int_\Omega\pi\, d\omega\right) =\int_\Omega d_L\pi\,d\omega .
		\end{equation*}
	\end{theorem}
	This property immediately derives of the writing of the integral into  $\mathcal{D}_0$. From the Stokes formula we deduce: 
	\begin{corollary}
		The Lie derivative  commutes with the exterior derivative of differential forms.
	\end{corollary}
	\begin{theorem} We have the properties:\\
		
		-\quad	If $s(\boldsymbol{x},t)$ is a scalar field moving with the fluid, then, the form field $\displaystyle\frac{\partial s}{\partial\boldsymbol{x}}$ is moving with the fluid.\\
		
		-\quad	If $\boldsymbol C(\boldsymbol{x},t)$ is a form field moving with the fluid, then the vector field\\ $\displaystyle\frac{1}{\rho}\,{\rm{curl}}\,\boldsymbol C^T$ is moving with the fluid.\\
		
		-\quad	If $\boldsymbol J(\boldsymbol x,t)$ is a vector field moving with the fluid, then the 3--form field isomorphic to the scalar field\,\ $\rm{div}(\rho\,\boldsymbol J) $ is moving with the fluid.\\
		
		-\quad	If $\alpha(\boldsymbol{x},t)$ and $\beta(\boldsymbol{x},t)$ are two scalar fields moving with the fluid, the 2--form field ${\partial \alpha}/{\partial\boldsymbol{x}}\wedge{\partial \beta}/{\partial\boldsymbol{x}}$ is moving with the fluid.
	\end{theorem}

	\subsection{Complements}
	\begin{theorem}
		$\rho$ being a scalar field  isomorphic to a 3-form,  $s$ a scalar field and $\boldsymbol{J}$ a vector field, all fields moving with the fluid (i.e. with a zero Lie's derivative with respect to the velocity field $\mathcal {U}$),   and  such that
		\begin{equation*}
		\frac{\partial s}{\partial \boldsymbol{x}}\, \boldsymbol{J}=0 \quad \text{and}\quad \text{div}\,(\rho\, \boldsymbol{J})=0,
		\end{equation*}
		then, there locally exists a scalar field $\eta$ moving with the fluid  and a mapping $(s, \eta) \longrightarrow f(s, \eta) $ such that 
		$$
		\rho\,\boldsymbol{J} = f(s,\eta)\  {\rm grad}\, s\wedge {\rm grad}\, \eta\quad   {\rm with}\quad \frac{d\eta}{dt} =0 .
		$$
	\end{theorem}
	{\bf Proof.\ } {\it
		$	\text{div}\,(\rho\, \boldsymbol{J})=0$ implies: \\
		
		Due to Clebsch's representation,	there locally exists 2 scalar fields $\tau$ and $\eta$ such that\,\  $\rho\, \boldsymbol{J} =
		\text{grad}\, \tau\wedge\text{grad}\, \eta$ \cite{Lamb}. The relation $\displaystyle\frac{\partial s}{\partial \boldsymbol{x}}\, \boldsymbol{J}=0$ proves that there  locally exists an application $F$ such that $F(s,\eta)=\tau$, and we obtain
		$\displaystyle\rho\, \boldsymbol{J} =\frac{\partial F}{\partial s}(s, \eta)\
		\text{grad}\, s\wedge\text{grad}\, \eta$. Vector field $\boldsymbol{J}$ moving with the fluid, $s$ and $\eta$ can be chosen moving with the fluid (i.e. have a zero Lie's derivative with respect to the velocity field $\mathcal U$).}
	\\
	
	All the previous properties are associated with conservation laws. The most classical ones are those of  conservation of mass, momentum and energy for conservative fluids which are represented by divergence forms in space-time $\mathcal W$, i.e. of the form \cite{Ruggeri}:
	\begin{equation*}
	\frac{\partial \boldsymbol a(x,t)}{\partial t} +\text{div}\left(\boldsymbol A(x,t)\right) =\boldsymbol 0,
	\end{equation*}
	where $\boldsymbol a(x,t)$ is a scalar field  and $\boldsymbol A(x,t)$ a matrix field of $\mathcal D_t$, respectively.\\
	To these conservation forms, it is natural to add non-divergence representations associated with  tensors with covariant order $p$, contravariant order $q$ and    zero Lie derivatives. To these tensors one can  associate scalars or 3-forms which allow to write conservation laws still in divergence forms. It is a way   to obtain universal relations in continuum mechanics \cite{Saccomandi}.  
	As example, we obtain:
	\\ 
	
	-\quad If $\boldsymbol J(\boldsymbol x,t)$ and   $\boldsymbol C(\boldsymbol x,t)$ have zero Lie derivatives, $\boldsymbol C(\boldsymbol x,t)\,\boldsymbol J(\boldsymbol x,t)$ is a scalar field moving with the fluid.\\
	
	-\quad If $\rho(\boldsymbol x,t)$ is a   3--form field moving with the fluid, $\rho(\boldsymbol x,t)\,\boldsymbol C(\boldsymbol x,t)\, \boldsymbol J(\boldsymbol x,t)$ is a   3-form field moving with the fluid.\\
	
	-\quad  It is the same for the field of mixed tensors  $\boldsymbol J(\boldsymbol x,t)\, \boldsymbol C(\boldsymbol x,t)$ 
	and we deduce that $\text{det}\left\{\boldsymbol J(\boldsymbol x,t)\, \boldsymbol C(\boldsymbol x,t) \right\}$ is a  scalar field moving with the fluid.\\

	-\quad It is similar for the vector field $\rho(\boldsymbol x,t)\; \boldsymbol W(\boldsymbol x,t)$ and the scalar field, \\ $\rho(\boldsymbol x,t)\; \boldsymbol C(\boldsymbol x,t)\;\boldsymbol W(\boldsymbol x,t)$, etc. \\
	
	-\quad  For all scalar field $\beta$ moving with the fluid, we get a conservation law in divergence form:
	\begin{equation*}
	\frac{\partial \rho\,\beta}{\partial t}+\text{div}(\rho\,\beta\,\boldsymbol u)\equiv \text{Div}(\rho\, \beta\,\mathcal U) =0 
	\end{equation*}
	and
	from all these conservation laws we can deduce first integrals along the flow,  and Kelvin's type theorems.
	\section{Other applications}
	
	\subsection{Electrodynamics of fluids}
	In the case of perfect non-conducting fluid motion (infinite electrical resistance), the equations of the electric field $\boldsymbol D$ and electric induction $\boldsymbol E$ are \cite{Landau,Lighthill}:
	\begin{equation*}
	{\rm div}\, \boldsymbol D = q\qquad {\rm and}\qquad 	{\rm curl}\,\boldsymbol E = \boldsymbol{0},
	\end{equation*}
	where $q$ is the volumetric electric charge. $\boldsymbol D$ and $\boldsymbol E$ are related by the relation $\boldsymbol D =\varepsilon \boldsymbol E$, where $\varepsilon$ is the electrical permeability of the medium. It depends on the density and   specific entropy.
	Our medium being supposed isotropic, the permeability tensor of the dielectric is scalar.
	The charge is conserved in the motion, i.e. the field $q$ is an isomorphic 3-form field moving with the fluid:
	\begin{equation*}
	\exists \left\{\boldsymbol X \in \mathcal D_0\longrightarrow q_0(\boldsymbol X)\right\}\,\ {\rm such\ that}\quad q\,	{\rm det}\,\boldsymbol F= q_0(\boldsymbol X).
	\end{equation*}
	Consequently ${\rm Div}\, (q\, \mathcal U) =0$ and due to the fact that ${\rm div}\, \boldsymbol D$ is a 3-form moving with the fluid, relation $(\rm det \boldsymbol F)\,div\, \boldsymbol D= div_0 \left({\rm det (\boldsymbol F) }\, \boldsymbol F^{-1}\boldsymbol D\right)$ yields:
	\begin{equation*}
	\boldsymbol D =\frac{\boldsymbol{F}}{ {\rm det}\, \boldsymbol{F}} \, \boldsymbol D_0(t,\boldsymbol x).
	\end{equation*}
	The field $\boldsymbol D_0$ is not necessarily a 2-form moving with the fluid. This is nevertheless the case when the fluid is in adiabatic motion.
	\subsection{Magnetodynamics of fluids}
	In the case of the adiabatic motion of a perfect fluid with zero electrical resistance, the equations verified by the magnetic field $\boldsymbol H(t,\boldsymbol x)$ are \cite{Landau}:
	\begin{equation*}
	{\rm div} \boldsymbol H =0 \quad {\rm and}\quad \frac{\partial\boldsymbol H} {\partial t}-{\rm curl} \left( \boldsymbol x\wedge \boldsymbol H\right)= \boldsymbol  0.
	\end{equation*}
	These relationships involve:
	\begin{equation*}
	\frac{d\boldsymbol H}{dt}+\boldsymbol H\, {\rm div}\, \boldsymbol u  -\frac{\partial\boldsymbol u}{\partial \boldsymbol x}\, \boldsymbol H =\boldsymbol 0  .
	\end{equation*}
	The magnetic field is such that the 2-form ${\rm det}(\boldsymbol H,.,.)$ is convected by the flow and it exists $\boldsymbol{H}_0(\boldsymbol{X})$ such that:
	\begin{equation*}
	\boldsymbol H(\boldsymbol z)= \frac{\boldsymbol{F}}{{\rm det}\, \boldsymbol{F}}\, \boldsymbol{H}_0(\boldsymbol{X}).
	\end{equation*}
	\\
	
	\section{Conclusion}
			
			Various concepts, such as the particle derivative, the convective derivative, the Jaumann derivative, etc., allow to express conservative fields frozen in a moving fluid medium and whose properties are given  by means of scalars and simple or multiple integrals.  
			A more natural method, linked to the definition of the motion, is that of the Lie derivative associated with a one-parameter group, corresponding to the diffeomorphism representing the motion of a fluid in space--time.  
			Its essential advantage is that it can express the variance of specific quantities defined on the reference space in Lagrangian variables. These quantities will have the structure of covariant order and contravariant order associated with scalars, vectors, shapes or tensors. Quantities depending only on the trajectory are moving with the fluid. They allow to express, as by Noether's theorem, some first integrals and conservation laws.

		\end{document}